\documentclass[conference]{IEEEtran}
\ifCLASSINFOpdf
 \usepackage[pdftex]{graphicx}
\else
\fi
\usepackage{amsmath}
\usepackage{amssymb}
\usepackage{algorithmic}

\usepackage{array}

\ifCLASSOPTIONcompsoc
  \usepackage[caption=false,font=normalsize,labelfont=sf,textfont=sf]{subfig}
\else
  \usepackage[caption=false,font=footnotesize]{subfig}
\fi
\usepackage{fixltx2e}
\usepackage{url}

\begin{document}
%
\title{Emotion Classification from Noisy Speech - A Deep Learning Approach}


\author{
\IEEEauthorblockN{Rajib Rana}
\IEEEauthorblockA{Institute for Resilient Regions (IRR)\\
University of Southern Queensland, Australia\\
rajib.rana@usq.edu.au}\\
\IEEEauthorblockN{Jeffrey Soar }
\IEEEauthorblockA{Faculty of Law and Enterprise\\
University of Southern Queensland, Australia\\
jeffrey.soar@usq.edu.au}
\and
\IEEEauthorblockN{Roland Goecke}
\IEEEauthorblockA{Institute for Resilient Regions (IRR)\\
University of Southern Queensland, Australia\\
rajib.rana@usq.edu.au}\\
\IEEEauthorblockN{Michael Breakspear}
\IEEEauthorblockA{Systems Neuroscience Group\\
Queensland Institute of Medical Research\\
michael.breakspear@qimrberghofer.edu.au}\\
\IEEEauthorblockN{Roland Goecke}
\IEEEauthorblockA{Institute for Resilient Regions (IRR)\\
University of Southern Queensland, Australia\\
rajib.rana@usq.edu.au}\\
\IEEEauthorblockN{Michael Breakspear}
\IEEEauthorblockA{Systems Neuroscience Group\\
Queensland Institute of Medical Research\\
michael.breakspear@qimrberghofer.edu.au}\\
}

%
\author{\IEEEauthorblockN{Rajib Rana\IEEEauthorrefmark{1},
Raja Jurdak\IEEEauthorrefmark{2},
Xue Li\IEEEauthorrefmark{3}, 
Jeffrey Soar\IEEEauthorrefmark{4} 
Roland Goecke\IEEEauthorrefmark{5}
Julien Epps\IEEEauthorrefmark{6} and
Michael Breakspear\IEEEauthorrefmark{7}
}
\IEEEauthorblockA{\IEEEauthorrefmark{1}Institute of Resilient Region (IRR)\\
University of Southern Queensland,
Springfield, QLD 4300\\ Email: rajib.rana@usq.edu.au}
\IEEEauthorblockA{\IEEEauthorrefmark{2}Sensor Systems Group, CSIRO-Data61\\
Email: raja.jurdak@csiro.au}
\IEEEauthorblockA{\IEEEauthorrefmark{3}School of Information Technology and Electrical Engineering, University of Queensland\\
xueli@itee.uq.edu.au}
\IEEEauthorblockA{\IEEEauthorrefmark{4}School of Management and Enterprise, University of Southern Queensland\\
jeffrey.soar@usq.edu.au}
\IEEEauthorblockA{\IEEEauthorrefmark{5}Information Technology \& Engineering, University of Canberra\\
roland.goecke@canberra.edu.au}
\IEEEauthorblockA{\IEEEauthorrefmark{6}School of Electrical Engineering and Telecommunications, University of New South Wales\\
j.epps@unsw.edu.au}
\IEEEauthorblockA{\IEEEauthorrefmark{7}Systems Neuroscience Group, Queensland Institute of Medical Research\\
Michael.Breakspear@qimrberghofer.edu.au}
}


\maketitle

\begin{abstract}
This paper investigates the performance of ``Deep Learning'' for speech emotion classification when the speech is compounded with noise. It reports on the classification accuracy and concludes with the future directions for achieving greater robustness for emotion recognition from noisy speech.
\end{abstract}


%
\IEEEpeerreviewmaketitle

\begin{figure*}[ht]
\centering
\includegraphics[width=1\linewidth]{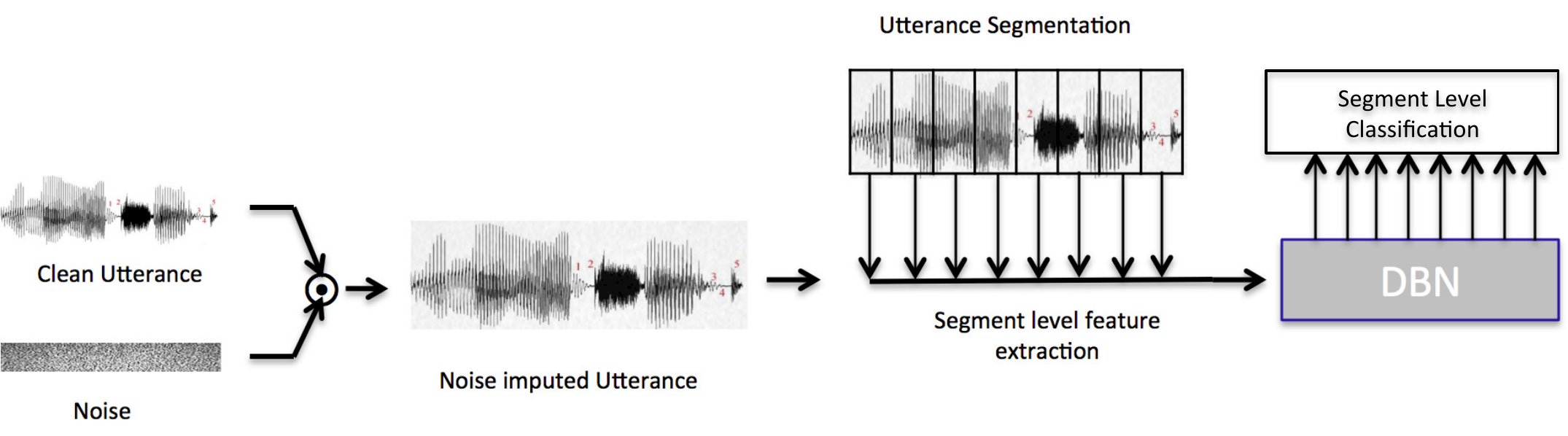}
\caption{Experimental setting for emotion clasiifcaiton from noisy speech.}
\label{fig:fig1}
\end{figure*}
\section{Introduction}

Deep Learning has revolutionised many fields of science including pattern recognition and machine learning. Its tremendous power in learning the non-linear relationships between visible and hidden layers over many layers of Deep Neural Networks (DNNs) makes it suitable for learning the most powerful features. Likewise, Deep Learning has revolutionised the field of speech recognition. A Deep Learning approach called, ``Deep Speech''~\cite{DBLP:journals/corr/HannunCCCDEPSSCN14} has significantly outperformed the state-of-the-art commercial speech recognition systems, such as Google Speech API and Apple Dictation. It achieves a word error rate of 19.1\%, whereas the best commercial systems achieved 30.5\% error. We capitalise on this success and aim to develop a deep learning framework for classifying emotion in speech. 

Emotion classification from speech is a widely studied topic. Before Deep Learning, Gaussian Mixture Models jointly with Hidden Markov Model were the most popular method for emotion classification from speech. After the Deep Learning breakthrough in speech recognition, the landscape of emotion recognition from speech has started to change. A number of studies have emerged where Deep Learning is used for speech emotion recognition. Linlin Chao~\cite{chao2014improving} et al. use Autoencoders, which is the simplest form of DNN. It is a simple 3-layer neural network where output units are directly connected back to input units. Typically, the number of hidden units is much less than the number of visible (input/output) ones. As a result, when data is passed through the network, the input vector is first compressed (encoded) to "fit" in a smaller representation, and then reconstructed (decoded) back. The task of training is to minimise an error or reconstruction, i.e. find the most efficient compact representation (encoding) of the input data. Autoencoders are efficient in discriminating some data vectors in favour of others, however, they are not generative. That is they cannot generate new data, and therefore cannot generate as rich features as the advanced DNNs.

The other form of Deep Neural Networks is the Deep Belief Networks, which uses stacked Restricted Boltzmann Machines (RBMs) to form a deep architecture. RBM has a two layer construction: a hidden layer and a visible layer and the learning procedure consists of several steps of Gibbs sampling, that is, in propagation,   hidden layer is sampled given the visible layer and in reconstruction visible layer  is sampled given the hidden layers. The reconstruction and the propagation passes are repeated adjusting the weights to minimise reconstruction error. Unlike Autoencoder, RBM is a generative model. It can generate samples from learned hidden representations. 

A number of studies have used DBNs for emotion classification from voice. In~\cite{le2013emotion}, DBN is used in conjunction with Hidden Markov Model. This work provides insights into similarities and differences between speech and emotion recognition. Authors in~\cite{niu2014acoustic} use a similar DBN-HMM architecture. In~\cite{sanchez2014deep}, authors test DBNs for feature learning and classification and also in combination with other classifiers (while using DBNs for learning features only) like k-Nearest Neighbour (kNN), Support Vector Machine (SVM) and others, which are widely used for classification~\cite{wei2012distributed}. An interesting approach called  Optimized Multi-Channel Deep
Neural Network
(OMC-DNN) has been proposed in~\cite{stolar2014optimized}, which for speech
emotion recognition uses input features generated as simple 2D black and white images representing graphs of the MFCC coefficients.

Finally, Convolution Neural Networks (CNNs)~\cite{huang2014speech}, which is another deep architecture, has also been used for emotion recognition from speech. CNNs are somewhat similar to RBMs, but instead of learning single global weight matrix between two layers, they aim to find a set of locally connected neurons, i.e., neurons that are spatially close to each other. CNNs are mostly used in image recognition. This is because, when dealing with high-dimensional inputs such as images, it is impractical to connect neurons to all neurons in the previous volume because such network architecture does not take the spatial structure of the data into account. Convolutional networks exploit spatially local correlation by enforcing a local connectivity pattern between neurons of adjacent layers: each neuron is connected to only a small region of the input volume. For the same reason, CNNs are hardly applicable for input other than images. Furthermore, this paper does not consider noisy speech.

Despite a number of attempts of using DBNs for emotion recognition from speech, no study has particularly looked into emotion recognition from ``noisy speech''. In this paper, we consider this challenge.

\section{Work in Progress} 
\begin{figure}[ht]
\centering
\includegraphics[width=1\linewidth]{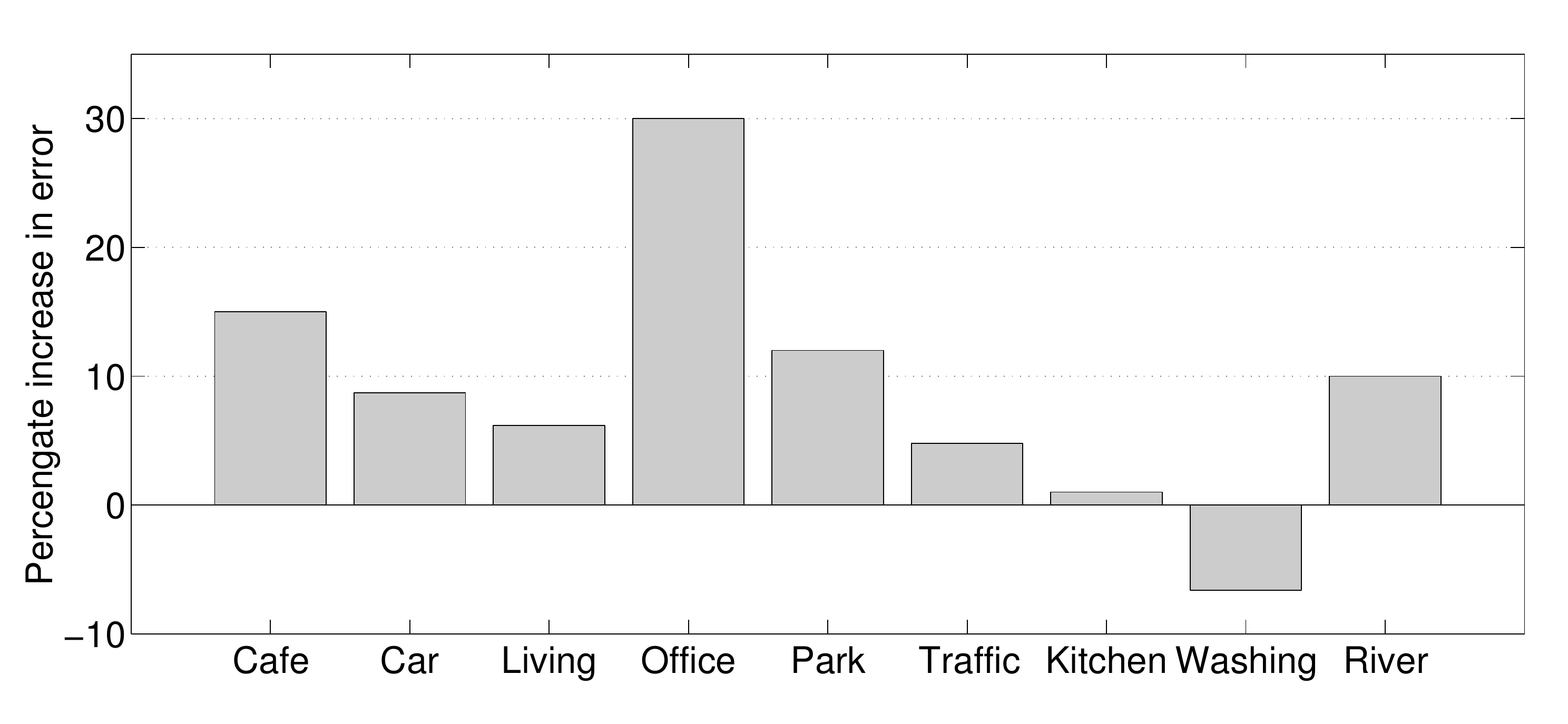}
\caption{Emotion recognition in various noisy conditions.}
\label{fig:fig2}
\end{figure}

Apart from the capacity of producing powerful discriminative features, DBNs are also naturally robust to noise. We use simulations to validate the robustness of DBNs. As shown in Figure~\ref{fig:fig1}, artificial noise is imputed to clean emotional utterance. Utterances are divided into segments and passed onto the DBNs for segment level emotion classification. 

The Berlin emotional speech database~\cite{burkhardt2005database} is used in experiments for classifying discrete emotions. In this database, ten actors (5m/5f) each uttered ten sentences (5 short and 5 longer, typically between 1.5 and 4 seconds) in German to simulate seven different emotions: anger, boredom, disgust, anxiety/fear, happiness, and sadness. Utterances scoring higher than 80\% emotion recognition rate in a subjective listening test are included in the database. We classify all the seven emotions in this work. The numbers of speech files for these emotion categories in the presented Berlin database are: anger (127), boredom (81), disgust (46), fear (69), joy (71), neutral (79) and sadness (62).
In order to simulate noise-corrupted speech signals, the DEMAND noise database~\cite{thiemann2013diverse} has been used in this paper. This database involves 18 types of noises including white noise and noises at the cafeteria, car, restaurant, etc. The DBN implementation has been adopted from~\cite{keyvanrad2014brief}. The DBN used in the paper uses three RBMs, where the first two RBMs use 1000 hidden unit each, and the third RBM uses 2000 hidden units. For each speech segment, a 13 coefficient MFCC vector~\footnote{MFCC features are widely used for acoustic signals~\cite{wei2013real}} is generated and used as the input to the DBN. The output is the classification result.  

Emotion recognition performance under different noise conditions is shown in Figure~\ref{fig:fig2}. To provide emphasis on robustness, we do not group the results based on the type of emotions, we rather present the overall classification accuracy. For each noise category, the results show the ``percentage difference in accuracy'' between emotion classification from clean and noisy speech. We made a number of observations:
\begin{enumerate}
\item The percentage errors can be categorised into three groups: error less than 10\%, error less than 20\% but greater than 10\% percent, error less than 30\% but greater than 20\%. 
\item From listening, the noise types causing the smallest error (error less than 10\%) have low magnitude and low variability, such as noises produced inside a car, inside kitchen etc.  Noises causing the higher errors have relatively higher variabilities. 
\item In general, accuracy is diminished in the presence of noise except in the case of ``washing'', when the accuracy increased. It is not unnatural for DBN to perform better in the presence of noise as this helps avoid overfitting, when the noise is not dominant. 
\end{enumerate}

\section{Conclusion and Future Work} 
We are currently developing methods to achieve greater robustness with DBNs for speech emotion recognition. This involves designing algorithms to obtain the optimal configuration (e.g., number of layers, nodes per layer etc.) of the DBNs. We are also developing an algorithm for optimal feature selection to achieve the best classification performance. Finally, we are conducting some in-depth analysis on the effect of various noise types in speech emotion recognition and how to overcome those effects.  

\bibliographystyle{IEEEtran}
\bibliography{reference.bib}
%
%
%
%

\end{document}